\documentclass[twocolumn,showpacs,prl,floatfix,aps]{revtex4}
\usepackage{amsmath}
\usepackage{hyperref}
\usepackage{graphicx}
\begin{document}
\title{Dynamics of Social Diversity}
\author{E.~Ben-Naim}\email{ebn@lanl.gov}
\author{S.~Redner}\email{redner@cnls.lanl.gov}
\altaffiliation{Permanent address:
Department of Physics, Boston University, Boston, Massachusetts,
02215 USA} \affiliation{Theoretical Division and Center for
Nonlinear Studies, Los Alamos National Laboratory, Los Alamos, New
Mexico 87545}
\begin{abstract}

We introduce and solve analytically a model for the development of
  disparate social classes in a competitive population.  Individuals
  advance their fitness by competing against those in lower classes,
  and in parallel, individuals decline due to inactivity.  We find a
  phase transition from a homogeneous, single-class society to a
  hierarchical, multi-class society.  In the latter case, a finite
  fraction of the population still belongs to the lower class, and the
  rest of the population is in the middle class, on top of which lies
  a tiny upper class.  While the lower class is static and poor, the
  middle class is upwardly mobile.

\end{abstract}
\pacs{87.23.Ge, 02.50.Ey, 05.40.-a, 89.65.Ef}
\maketitle

A distinguishing feature of developed societies is the existence of
social classes.  What accounts for this diversity?  Do individuals
advance their position as a result of innate talent, inherited wealth,
or plain luck?  Our aim is to construct a minimalist interacting agent
model that accounts for the development of social diversity.

The phenomenon of social diversity has been extensively investigated
both in the social \cite{C,G} and the biological sciences \cite{L,W}.
Social hierarchies have been observed empirically in a wide range of
animal populations, from insects \cite{W1} to mammals \cite{A}, and to
primates \cite{VS} and humans \cite{C1}.  An appealing route for
modeling social dynamics is to use interacting agent systems
\cite{vfh,ckfl}. Quantitative \cite{ww} and large scale simulations
\cite{bes} of physics-inspired interacting particle systems have been
used to model observed macroscopic collective phenomena using
microscopic agent interactions \cite{wdan,bkr,smo}.

Motivated by empirical observations of bumblebee communities
\cite{VH}, Bonabeau et al.\ recently introduced an agent-based model
of social diversity \cite{BTD} where each individual is endowed with a
fitness-like variable that evolves by two opposing processes.  The
first is competition: when two agents interact, one individual becomes
more fit (gains status) and the other becomes less fit, with the
initially fitter individual being more likely to win.
Counterbalancing the competition, the winning probability for the
fitter agent decreases as the time from the last competition
increases.  Investigations of this model have found a transition from
a homogeneous to a hierarchical society as the relative strengths of
these two processes are varied \cite{BTD,SS,MSK}.

In this letter, we introduce a simplified version of the Bonabeau
social diversity model that accounts for the competing processes of
advancement by competition and decline by inactivity via a single
parameter.  By solving the underlying rate equations, we find a phase
transition from a homogeneous, single-class society to a hierarchical,
multi-class society.  In the latter phase, the lower class is
destitute and static while the middle class is dynamic and has a
continuous upward mobility.

In our model, an agent is endowed with an integer fitness value $k\geq
0$.  All agents start with zero fitness and the fitness changes by two
processes: (i) advancement by competition and (ii) decline by
inactivity.  In the competition step, when two agents interact, their
fitness changes according to
\begin{equation}
\label{up}
(k,j)\to (k+1,j),
\end{equation}
for $k\geq j$.  When two equally fit agents compete, only one
advances.  Without loss of generality, the rate of this process is set
to one.  We also consider the mean-field limit where any pair of
agents is equally likely to interact.  The rationale behind this
``rich gets richer'' dynamics is obvious: fitter individuals are
better suited for, and hence benefit from, competition.  When decline
occurs, individual fitness decreases as
\begin{equation}
\label{down}
k\to k-1
\end{equation}
with a rate $r$.  This process reflects the natural tendency for
social status to decrease in the absence of interactions.  The lower
limit for the fitness is $k=0$; once an individual reaches zero
fitness, there is no further decline.  The model is characterized by a
single parameter, the rate of decline $r$.

Let $f_k(t)$ be the fraction of agents with fitness $k$ at time $t$. This
distribution obeys the nonlinear master equation
\begin{eqnarray}
\label{dis-eq}
\frac{d f_k}{dt}&=&r(f_{k+1}-f_k)+f_{k-1}F_{k-1}-f_kF_k
\end{eqnarray}
for $k\geq 0$, where $F_k=\sum_{j=0}^k f_j$ is the cumulative
distribution. The boundary condition is $f_{-1}=0$ so that
$df_0/dt=rf_1-f_0^2$, and the initial condition is
\hbox{$f_k(0)=\delta_{k,0}$}. The first two terms in
Eq.~(\ref{dis-eq}) account for the decline of the fitness of an agent,
while the last two terms account for advancement \cite{bkm}.

To understand the behavior of this system, we focus on the cumulative
distribution $F_k$, from which the individual densities are
\hbox{$f_k=F_k-F_{k-1}$}.  From the master equation (\ref{dis-eq}),
the cumulative distribution satisfies
\begin{equation}
\label{cum-eq}
\frac{dF_k}{dt}=r(F_{k+1}-F_k)+F_k(F_{k-1}-F_k),
\end{equation}
for $k\geq 0$. The boundary condition is $F_{-1}=0$ so that
$dF_0/dt=r(F_1-F_0)-F_0^2$, and the initial condition is $F_k(0)=1$.

\smallskip\noindent{\bf Homogeneous vs.\ Hierarchical Societies.}  Our
social diversity model undergoes a phase transition from a homogeneous
to a hierarchical society. This transition follows from the continuum
limit of the master equation (\ref{cum-eq}) for the cumulative
distribution
\begin{equation}
\label{cum-cont}
\frac{\partial F}{\partial t}=(r-F)\frac{\partial F}{\partial k}.
\end{equation}
For finite fitness, the cumulative distribution approaches a steady
state in the long-time limit. Then either $F=r$ or $\partial
F/\partial k=0$. Invoking the bound $F\leq 1$, we conclude that either
$F=r$ or $F=1$. Therefore, $L$, the fraction of the population with
finite fitness exhibits a phase transition
\begin{equation}
\label{mass}
L=
\begin{cases}
r&r< 1;\\
1&r\geq 1.
\end{cases}
\end{equation}
When competition is weak, the entire population has a finite fitness,
while for strong competition, only a fraction $L<1$ of the population
has a finite fitness.

\begin{figure}[t]
 \vspace*{0.cm}
\includegraphics*[width=0.45\textwidth]{fig1.eps}
\vskip -1.33in\hskip 1.0in\includegraphics*[width=0.275\textwidth]
{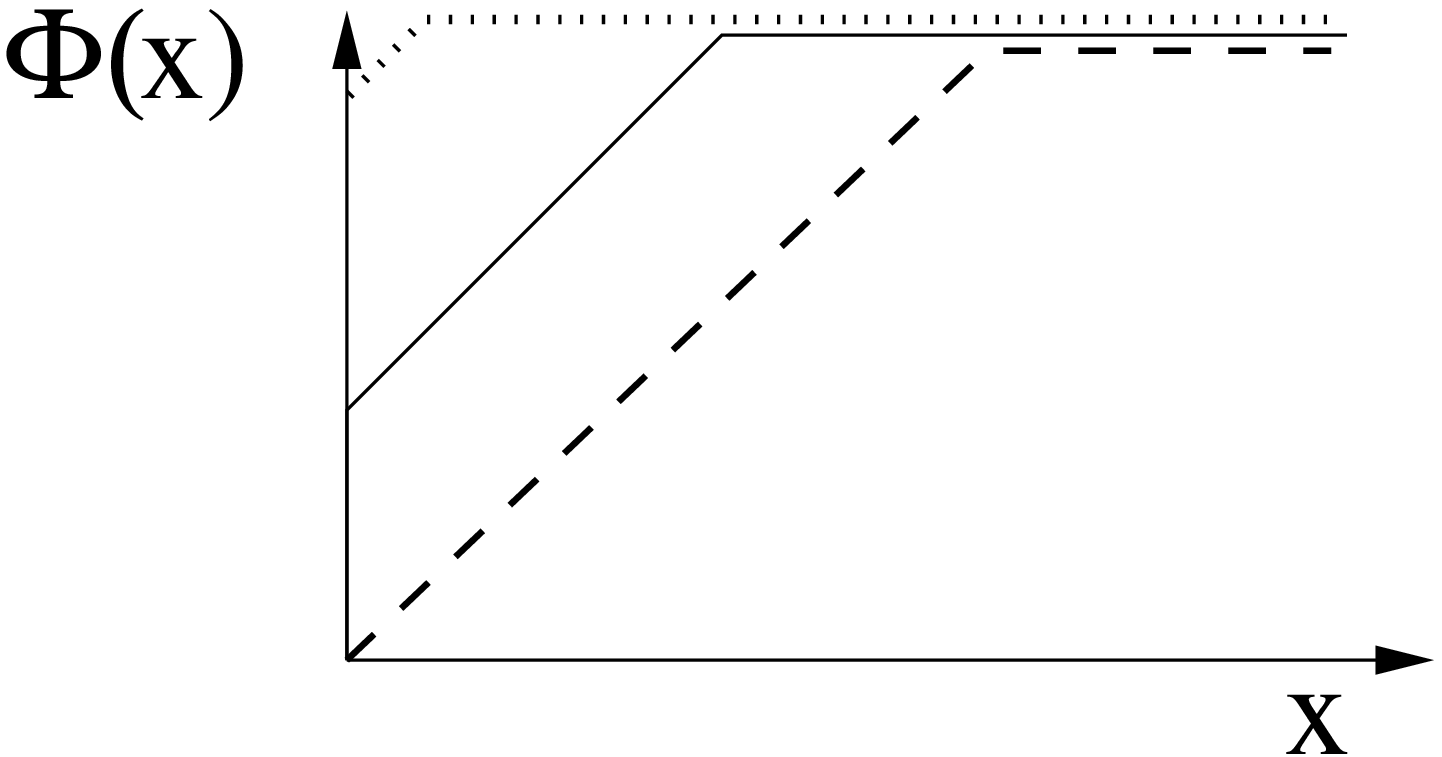} \vskip 0.25in \caption{The middle class. The scaled
cumulative distribution $\Phi(x)$ versus $x$ for $r=1/2$ at $t=250$
(dotted), $1000$ (dashed), $4000$ (dot-dashed). The solid line is the
theoretical prediction (\ref{cum-sol}).  The inset shows the
qualitative behavior for $r=0$ (dashed), $r\approx 1/2$ (solid), and
$r\approx 1$ (dotted). \label{middle}}
\end{figure}

We shall see that the quantity $L$ is the size of the lower class,
while the complementary fraction $1-L$ is the size of the middle
class, whose fitness increases indefinitely.  Thus for $r\geq 1$, the
society is homogeneous and consists of a single lower class.  However
for $r<1$, there is a hierarchical society that contains a distinct
lower class, and a distinct a middle class.  When $r=0$, the lower
class disappears entirely.

\smallskip\noindent{\bf Middle Class Dynamics.}  The picture presented
above is confirmed by analyzing the dynamics of the middle class.
Applying dimensional analysis to the governing Eq.~(\ref{cum-eq})
suggests that the characteristic fitness of the middle class increases
linearly with time, $k\sim t$. Thus, we posit the scaling form
\begin{equation}
\label{scaling}
F_k\simeq \Phi(k/t)
\end{equation}
with the boundary condition $\Phi(\infty)=1$.  Substituting
Eq.~(\ref{scaling}) into (\ref{cum-cont}), the scaling function
satisfies \hbox{$x\,d\Phi/dx=(\Phi-r)\,d\Phi/dx$} where $x=k/t$. The solution is
either \hbox{$\Phi(x)=r+x$} or $d\Phi/dx=0$.  As a result
(Fig.~\ref{middle})
\begin{equation}
\label{cum-sol}
\Phi(x)=
\begin{cases}
r+x&x< 1-r;\\
1&x\geq 1-r.
\end{cases}
\end{equation}
Remarkably, the scaling function for the cumulative distribution is
piecewise linear and thus non-analytic.

\begin{figure}[t]
 \vspace*{0.cm}
\includegraphics*[width=0.475\textwidth]{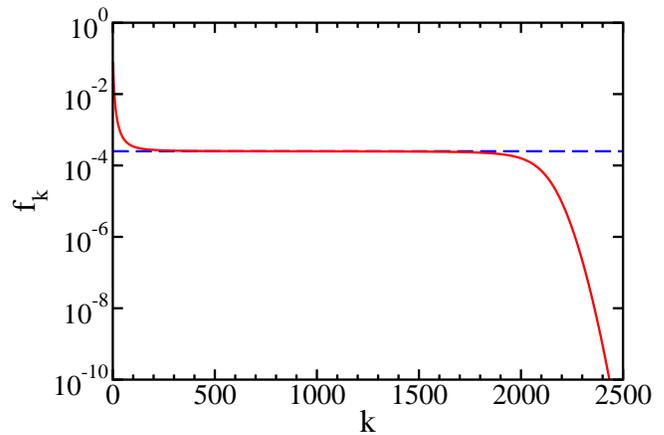}
\caption{The lower, middle, and upper classes. The fitness
  distribution $f_k$ versus $k$ for $r=1/2$ at $t=4000$ (solid line),
  showing the lower \hbox{($k\alt 60$)}, middle, and upper
  \hbox{($k\agt 2000$)} classes.  Also shown for reference is the
  plateau $f_k=1/t$ (dashed line).  The distribution decays as $1/k^2$
  in the lower class up to the diffusive scale $k_{\rm lower}\sim
  (2rt)^{1/2}$.  The distribution is constant in the middle class up
  to a ballistic scale $k_{\rm upper}=(1-r)t$, beyond which there is
  an upper class that has a Gaussian decay.
  \label{all}}
\end{figure}

The scaling function (\ref{cum-sol}) has a number of basic
implications. First, the quantity $\Phi(0)=r$ is the fraction of the
population that belongs to the lower class, confirming the prediction
of Eq.~(\ref{mass}). This behavior is reminiscent of a physical Bose
condensate, where a finite fraction of the population occupies the
zero fitness (in scaled units) ground state.  In this sense, the
entire lower class is destitute.  When there is only competition
($r=0$), the society consists of a continuously-improving middle
class.  In this case, a formal exact solution of the master equations
is possible \cite{unpub}.

We can alternatively write the fitness distribution in the scaling
form $f_k\simeq t^{-1}\phi(k/t)$.  The corresponding scaling function
is \hbox{$\phi(x)=d\Phi/dx=r\delta(x)+1$} for $x\leq 1-r$ and $\phi(x)=0$
otherwise.  The middle class thus has a {\em constant\/} fitness
distribution
\begin{equation}
\label{flat}
f_k\simeq t^{-1},
\end{equation}
for $k<k_{\rm upper}=(1-r)t$.  The lot of the middle class is constantly
improving, as the fitness extends over a growing range and the
average fitness increases linearly with time.

Numerical integration of the master equation confirms these
predictions (Figs.~\ref{middle} and \ref{all}).  We used a
fourth-order Adams-Bashforth method \cite{zwillinger} with accuracy to
$10^{-10}$ in the distribution $F_k$.  Our numerical data was obtained
by integrating $F_k$ for $0\leq k<20000$.

\smallskip\noindent{\bf Lower Class Dynamics.} The fitness of the
lower class is finite; in other words, the fitness distribution is in
a steady state. This distribution can be determined by setting the
time derivative in the rate equation to zero. Writing $F_k=L(1-G_k)$,
so that the deviation $G_k$ vanishes at large $k$, Eq.~(\ref{cum-eq})
gives
\begin{equation}
\label{recursion}
r\,\frac{G_{k+1}-G_k}{G_k-G_{k-1}}=L(1-G_k).
\end{equation}

\begin{figure}[t]
 \vspace*{0.cm}
\includegraphics*[width=0.475\textwidth]{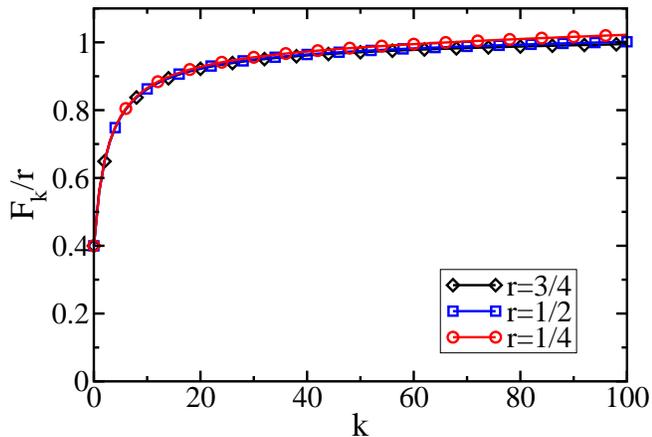}
\caption{The lower class.  The cumulative distribution, normalized by
  the rate $r$, $F_k/r$ is plotted versus $k$.  Shown are simulation
  results for $r=1/4$ (circles), $r=1/2$ (squares), $r=3/4$ (diamonds)
  at time $t=10^4$.
\label{lower}}
\end{figure}

The fitness distribution is fundamentally different in the two phases.
In the homogeneous society phase ($r\geq 1$ and $L=1$), the deviation
$G_k$ decays rapidly at large fitness.  Replacing the right-hand side
of Eq.~(\ref{recursion}) by $1$ for large $k$, the solution is simply
$G_k\sim r^{-k}$.  Therefore
\begin{equation}
\label{exponential}
f_k\sim r^{-k}.
\end{equation}
The fitness distribution decays exponentially, so that the lower class
is confined to a small range of fitness values.  The characteristic
fitness $1/\ln r$ diverges as the transition is approached.  The
society is homogeneous as it contains a single social class, the lower
class, that does not evolve with time.

In the hierarchical society phase, (where \hbox{$r< 1$} and $L=r$),
the fitness distribution is universal, as the recursion relation
(\ref{recursion}) becomes $r$-independent,
\hbox{$(G_{k+1}-G_k)/(G_k-G_{k-1})=1-G_k$}. This shows that $F_k/r$ is
a universal, $r$-independent distribution (Fig.~\ref{lower}). We start
by treating $k$ as a continuous variable, because the fitness range
becomes large as $r\downarrow 1$.  We thus expand the differences in
Eq.~(\ref{recursion}) to second order.  Since $G''\ll G'$, where prime
denotes differentiation with respect to $k$, we find $G''+GG'=0$.
Integrating once and invoking $G\to 0$ as $k\to \infty$, gives
$G'+\frac{1}{2}G^2=0$.  Asymptotically, $G\simeq 2k^{-1}$, and then by
using $f_k=F_k-F_{k-1}$, we find
\begin{equation}
\label{powerlaw}
f_k\simeq 2r\,k^{-2}.
\end{equation}
The lower class has a power-law fitness distribution with mean
fitness that diverges logarithmically in the upper limit.  While
the lower class is still static, it is not as destitute as in the
homogeneous society phase.

The transition between the lower and middle class occurs when
$2r/k^{2}\approx 1/t$, {\it i.e.}, where the power-law distribution
(\ref{powerlaw}) matches the uniform distribution (\ref{flat}).
Consequently, the lower class is confined to a diffusive boundary
layer of thickness
\begin{equation}
\label{diffusive}
k_{\rm lower}\sim (2rt)^{1/2}.
\end{equation}
Beyond this diffusive scale, lies the middle class whose constant
density (\ref{flat}) extends over the range \hbox{$k_{\rm lower}< k<
k_{\rm upper}$}. In the hierarchical society phase, the fitness
distribution consists of the stationary component (\ref{powerlaw})
that defines the lower class and the evolving component
(\ref{scaling}) that defines the middle class. The extent of the
stationary region indefinitely grows with time.

\begin{figure}[t]
 \vspace*{0.cm}
\includegraphics*[width=0.475\textwidth]{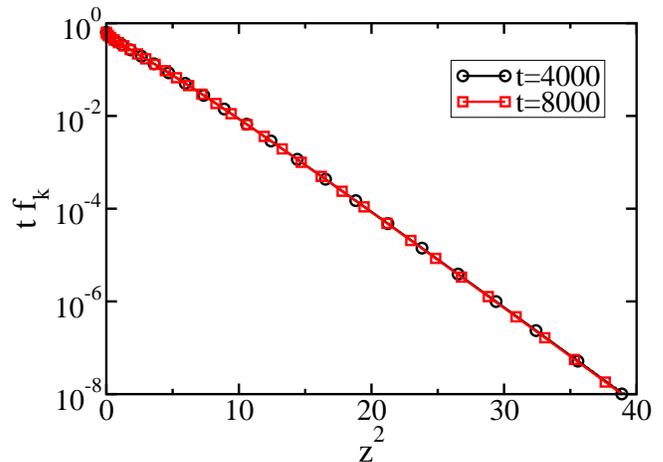}
\caption{The upper class. Shown is the normalized tail of the fitness
  distribution: $tf_k$ versus $z^2$, with the scaling variable
  $z=(k-vt)/\sqrt{Dt}$, for $r=1/2$ at times $t=4000$ (circles) and
  $t=8000$ (squares).
\label{upper}}
\end{figure}

We thus conclude that the lower class is always static, being in a
steady-state independent of the rate of decline $r$.  In a homogeneous
society, the lower class has an exponentially decaying fitness
distribution that lies within a narrow fitness range.  In a
hierarchical society, the lower class fitness distribution decays
algebraically and its range grows diffusively with time.

\smallskip\noindent{\bf Upper Class Dynamics.}  The upper class is
defined by the subpopulation whose fitness lies beyond \hbox{$k_{\rm
upper}=(1-r)t$}.  We probe the tail of this fitness distribution by
again considering the deviation $G_k$,  defined by
$F_k=1-G_k$.  This deviation obeys the convection-diffusion equation
\begin{equation}
\frac{\partial G_k}{\partial t}+v\frac{\partial G_k}{\partial k}=
D\frac{\partial^2 G_k}{\partial k^2}
\end{equation}
with upward drift velocity $v=(1-r)$ and diffusion coefficient
$D=(1+r)/2$.  The boundary condition $G(k=vt)\propto t^{-1}$ is set by
matching the density at the top of the middle class with that at the
bottom of the upper class. Consequently, the fitness distribution,
$f=-\partial G/\partial k$, follows the scaling form
(Fig.~\ref{upper})
\begin{equation}
\label{f-upper}
f_k(t)\simeq t^{-1}\psi\left(\frac{k-vt}{\sqrt{Dt}}\right),
\end{equation}
where the scaling function has the Gaussian tail \hbox{$\psi(z)\sim
\exp(-z^2/2)$}, as $z\to \infty$.

The upper class is thus confined to a diffusive boundary layer that
grows as $\sqrt{Dt}$.  From Eq.~(\ref{f-upper}), the upper class
contains a fraction $\propto 1/\sqrt{t}$ of the total population.
Finally, for a finite population of $N$ agents, we deduce from the
extreme statistics criterion, $Nf_k\sim 1$, that for the fittest agent
$k_{\rm extreme}\sim vt+\sqrt{t \ln N}$.

We comment that the deceptively simple master equation exhibits a
remarkable triple-deck structure, with a stationary component,
followed by two transient components.  Interestingly, the asymptotic
fitness distribution is described by a non-analytic scaling function.

In summary, we introduced a minimal model of social diversity in
which the two driving mechanisms are advancement by competition and
decline by inactivity.  An idealized but plausible social structure
emerges: either a homogeneous society with a single lower class, or a
hierarchical society with multiple classes. The lower class is always
static, while the middle class and the (tiny) upper classes are
upwardly mobile. In a hierarchical society, the lower and the upper
classes are confined to boundary layers that are much smaller than the
dominant scale that characterizes the fitness of the middle class.

There are numerous interesting questions suggested by this work.  For
example, what is the time history of an individual?  How rigid is the
social hierarchy and how does it depend on the population size?  What
happens if each individual is also endowed with an intrinsic fitness?
Last, does non-trivial spatial organization emerge when agents move
locally in space?

We thank D. Stauffer for stimulating our interest in social
 diversity, K. Kulakowski for an informative discussion, and D. Watts
 for literature advice.  We also acknowledge financial support from
 DOE grant W-7405-ENG-36 (EB and SR) and NSF grant DMR0227670 (SR).

\end{document}